\documentclass[
    ,final            
  ]
  {aipproc}

\layoutstyle{6x9}

\begin{document}

\title{Hadron Blind Detector for the PHENIX Experiment at RHIC}


\author{A. Milov for the PHENIX Collaboration}{
  address={Brookhaven National Laboratory, Upton, NY, 11973, USA}
}

\begin{abstract}
The PHENIX collaboration has designed a conceptually new Hadron Blind
Detector (HBD) for electron identification in high density hadron
environment. The HBD will identify low momentum electron-positron pairs to
reduce the combinatorial background in the mass region below 1
GeV/c$^{2}$. The HBD shall be installed in PHENIX during
the 2007 physics run.

The HBD is a windowless proximity focusing Cherenkov detector with a radiator length of 50 cm,
$CsI$ photocathode and three layers of Gas Electron Multipliers (GEM) for gas
amplification. Pure $CF_{4}$ serves both as a radiator and as a detector gas.
The radiation budget of the device is less than 3\% of a radiation length.
\end{abstract}

\maketitle

\section{Physics goal}
The study of low-mass electron-positron pairs in Heavy Ion Collisions
is a powerful tool to investigate properties of the newly discovered
strongly coupled Quark Gluon Plasma. Results obtained by the
NA45 and NA60 experiments at CERN~\cite{na45,na60} show an excess of
pairs produced in the mass region below 1GeV/c$^2$. The PHENIX
experiment at RHIC measured the low mass dilepton
region~\cite{phenix_dilept}, however a detailed study of dileptons is
very difficult without suppression of the combinatorial background 
coming from electron-positron pairs with small opening
angle. Their primary source is the decays
of $\pi^{0}$ mesons and $\gamma$-conversions. The main purpose of the PHENIX upgrade
with the HBD detector is the rejection of close $e^{+}e^{-}$ pairs by two
orders of magnitude.

\section{HBD Concept}
An extensive simulation done for the HBD~\cite{loi} demonstrated that
the only way to suppress the dilepton combinatorial background without loosing the
signal requires electron identification. The $e/\pi$ rejection
factor must be 100 with an electron detection efficiency of
90\%. The HBD must cover solid angle $\sim$20\% larger than the nominal PHENIX
acceptance. Detectors based on Cherenkov radiation are the most effective
choice to meet such requirements.

Cherenkov detector consists of a radiator in which particles shall
follow straight trajectories. This competes with the high resolution
tracking which benefits from the increase of the magnetic field. The PHENIX
detector~\cite{phenix} has two magnetic field coils which allow to increase the magnetic
field for better tracking or compensate it within the radius of 50-60
cm from the interaction point and keeping the initial angles of all tracks.

An important element of the classic Ring Imaging Cherenkov Detector is the
focusing mirrors located at the end of the radiator. In
PHENIX configuration the mirror cannot be used because of the
geometrical constrains. Therefore it was decided to replace
the mirrors with the light sensitive detectors placing them in the
path of all particles produced in the collisions. The HBD detection
unit thus must be sensitive to UV light and blind to all hadrons
traversing it. It must also keep its radiation budget below 3\% of the
radiation length and fit into $R\leq60$ cm.

\subsection{The photocathode and the gas}
A crucial decision for the Cherenkov counter is the choice of the
photo-sensor. Amid a variety of choices the only considered were those which
fit the radiation length budget, i.e. solid film and gaseous/aerosol
sensors. The second option, however, requires to separate radiator and
amplification into two different gaseous volumes. It significantly
increases the radiation length and besides that the window serves as
an auxiliary source of Cherenkov radiation. The first option is
preferable in spite of the same gas must be used as the Cherenkov radiator and
the electron multiplication media.

\begin{figure}
  \includegraphics[height=.2\textheight]{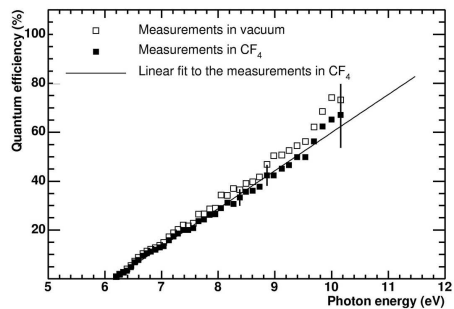}
  \caption{Quantum efficiency of the $CsI$ photocathode in the vacuum
(open symbols) and in the pure $CF_{4}$ atmosphere (full symbols).\label{fig:qe}}
\end{figure}

The most widely used film photocathode is the $CsI$ reaching 80\% quantum
efficiency (Q.E.) in extreme ultra-violet (E.U.V.) part of the spectra, see
fig.~\ref{fig:qe}. To match the photocathode Q.E. one has to use the gas which is
transparent in the E.U.V. That limits the choice of gases to mixtures of carbofluorides and noble
gases having the deepest E.U.V. cut-off wavelength. Among them the
pure $CF_{4}$ transparent up to 11.5 eV and with the refraction index
of n=1.00062\footnote{depends on the photon energy, given value
corresponds to visible light} is considered the primary choice. Based on the R\&D carried out during the development
stage~\cite{RandD} the electrons can be effectively extracted into
$CF_{4}$ atmosphere with up to 85\% efficeincy compare to vacuum (full symbols in
fig.~\ref{fig:qe}). The same parameter in pure $Ar$ of never exceeds 60\%.

\subsection{The amplification unit}
The main challenge is to build a detector sensitive to the smallest possible charge
produced by the Cherenkov photons and not sensitive to the much larger
ionization coming from hadrons traversing it. The difference between
the two
processes is in the localization of the primary charges. In the first
case they appear on the surface of the photocathode and in the second
inside the gas volume. The detector concept utilizing primary charge localization was suggested by Y.~Giomataris
and G.~Charpak~\cite{giomataris} in 1991. Electrons extracted from the
photocathode surface undergo full amplification in a parallel plate
detector whereas electrons produced inside the detector sensitive volume are
only partially amplified. A prototype of such detector was built and studied by the group of
T.~Hemmick at Stony Brook University in 1996~\cite{tom}. 

The major problem with this approach are the avalanche photons shining
back to the photocathode. Since the gas used in the detector is the same gas as used
in the radiator no admixtures can be used to quench the feedback. The advent of the new
detector, the Gas Electron Multiplier (GEM), invented by F.Sauli
in 1997~\cite{sauli} made it possible to overcome this difficulty. Among other features of the GEM is that the
photon feedback is blocked by geometry of the detector.

\begin{figure}
  \includegraphics[height=.2\textheight]{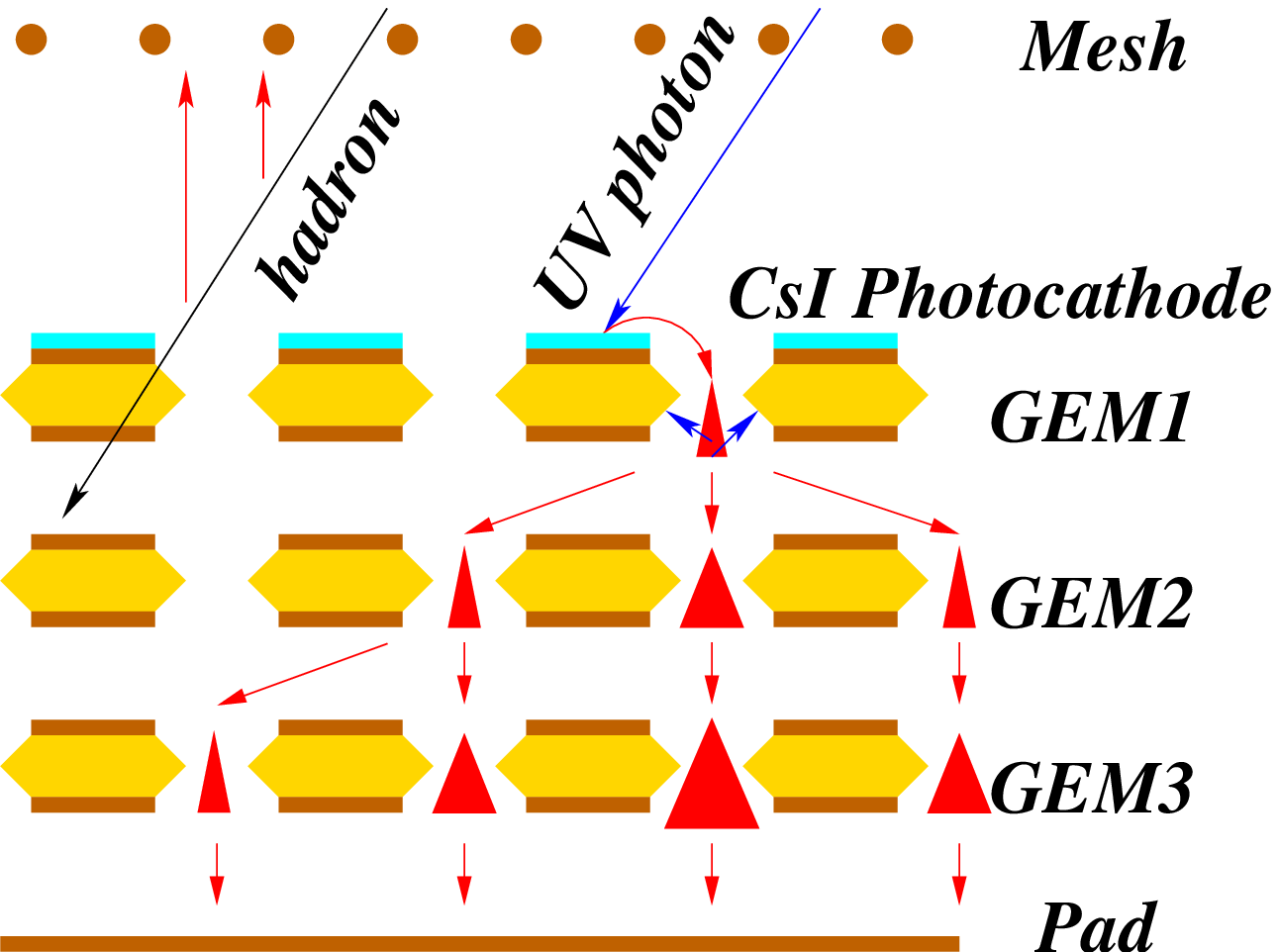} \hspace{1cm}
  \includegraphics[height=.25\textheight]{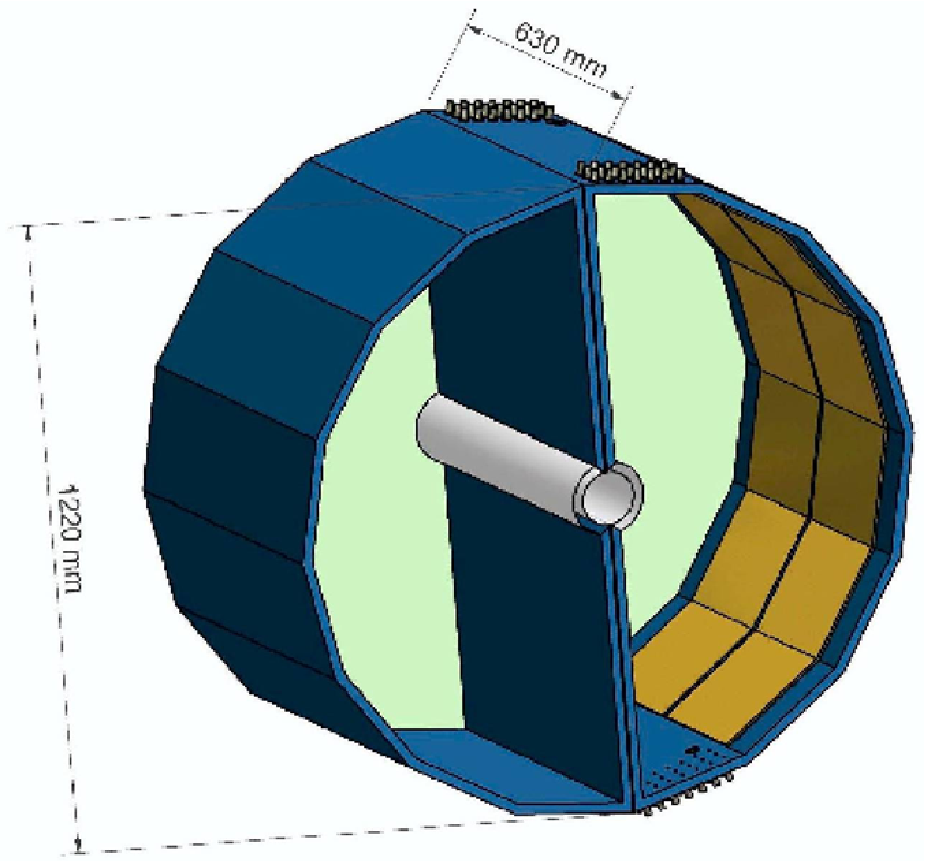}
  \caption{Left: Schematics of the amplification unit. Red and blue lines
are electrons and photon tracks respectively. Right: Schematics of the
final detector\label{fig:gems}}
\end{figure}

By evaporating a thin layer of $CsI$ onto the GEM
surface one can convert a GEM into highly efficient semi-transparent
photocathode\footnote{photoelectron extraction on the same side as
the incoming photons is several times(!) more efficient than pulling them through
the photosensitive layer. E.g.: a conventional PMT scheme}. It was demonstrated~\cite{RandD} that the electric field required
for electron amplification inside the GEM holes is sufficient to extract
electrons form any point of the surface of the GEM and direct them
into the nearest hole. Electrons produced by the
ionizing particles in the gas volume above the photocathode can be removed by an inversely
biased drift field making the detector insensitive to ionizing
radiation. This scheme is shown in fig~\ref{fig:gems}. 

Three GEMs are used in the detector to provide the gas amplification
of $10^{4}$~\cite{sasha}. Without a focusing mirror the Cherenkov photons on the detector
populate a circle not a ring thus the single electron detection does
not improve the pattern recognition. In the HBD the signal is collected by
hexagonal pads of a size slightly smaller than the size of the
circle~\footnote{typical area of the pad in the final detector is 6.3
$cm^{2}$}. This maximizes the signal collected on a single pad and provides extra rejection power to the
hadrons. Ionization is always localized in one single pad while primary
electrons producing Cherenkov light spread it over at least two adjacent pads.

The separation between a single electron and a close electron pair is
done by the analyzing the amplitudes. According to the
simulation the number of photoelectrons produced by a primary electron
is 36. Such signal can be reliably distinguished from twice that number
produced by a dilepton pair with a small opening angle.

\section{Detector layout}
The HBD detector
is shown in the right panel of fig.~\ref{fig:gems}~\cite{it}. The detector consists
of two half-cylinder volumes made of honeycomb
fiberglass sandwich panels. The top and bottom sectors, sitting outside
the PHENIX fiducial acceptance are used for
services and the six inner sectors are covered by a single piece Kapton film with
1152 pads printed on it. The film also serves as an additional gas
seal. Particles enter the detector trough a 0.12 mm
aluminized mylar window. 12 gas amplification modules as described
above, are located on the inner side of each HBD vessel. Each module is
27$\times$23 cm$^2$. GEMs used in them are subdivided into 28 HV modules to reduce
energy stored in them in case of electric discharge. All GEMs in a
module are powered with a single HV power supply  through a
resistive chain. The HBD will be operational
in PHENIX in 2007 physics run.

\begin{theacknowledgments}
Authors acknowledge support from the Department of Energy, NSF
(U.S.A.) and US-Israel BSF. Work of the speaker is supported by the
Goldhaber Fellowship at BNL with funds provided by Brookhaven
Science Associates. 
\end{theacknowledgments}

\end{document}